\documentclass[prl,twocolumn,amsmath,amssymb,superscriptaddress]{revtex4-2}

\usepackage{graphicx}
\usepackage{dcolumn}
\usepackage{bm}

\begin{document}

\title{Sliding Charge Density Wave Observed through Band Structure}


\author{S. Mandal}
\affiliation{Laboratoire de Physique des Solides, Université Paris-Saclay,CNRS, 91405 Orsay Cedex, France.}
\author{D. Ghoneim}
\affiliation{Laboratoire de Physique des Solides, Université Paris-Saclay,CNRS, 91405 Orsay Cedex, France.}
\affiliation{European X-ray Free-Electron Laser Facility, Holzkoppel 4, 22869 Schenefeld, Germany.}
\author{A.A.Sinchenko}
\affiliation{Laboratoire de Physique des Solides, Université Paris-Saclay,CNRS, 91405 Orsay Cedex, France.}
\author{V.L.R.Jacques}
\affiliation{Laboratoire de Physique des Solides, Université Paris-Saclay,CNRS, 91405 Orsay Cedex, France.}
\author{K. Wang}
\affiliation{Laboratoire de Physique des Solides, Université Paris-Saclay,CNRS, 91405 Orsay Cedex, France.}
\author{L. Ortega}
\affiliation{Laboratoire de Physique des Solides, Université Paris-Saclay,CNRS, 91405 Orsay Cedex, France.}
\author{J. \'Avila}
\affiliation{Synchrotron SOLEIL, L’Orme des Merisiers, 91190 Saint-Aubin, France}
\author{P. Dudin}
\affiliation{Synchrotron SOLEIL, L’Orme des Merisiers, 91190 Saint-Aubin, France}
\author{A. Tejeda$^*$}
\affiliation{Laboratoire de Physique des Solides, Université Paris-Saclay,CNRS, 91405 Orsay Cedex, France.}
\author{D. Le Bolloc’h$^*$}
\affiliation{Laboratoire de Physique des Solides, Université Paris-Saclay,CNRS, 91405 Orsay Cedex, France.}

\begin{abstract}
An incommensurate CDW may have the ability to slide, i.e., to generate an excess of current when the system is submitted to an external field. Sliding phenomenon is closely related to deformation of the periodic lattice distortion associated to the CDW. In principle, however, the sliding state can also be observed through the band structure. Here we show that broken symmetry in k-space is observed by angle-resolved photoemission spectroscopy (ARPES) in the sliding regime of TbTe$_3$, which could be consistent with theoretical predictions.
\end{abstract}

\maketitle

The Charge Density Wave (CDW) phase is a broken symmetry state widely occurring in condensed matter and coexisting with several other phases, including superconductivity\cite{Ghiringhelli2012}. The electronic instability involved in this transition is induced by electron-phonon coupling which leads to two consequences. First a gap opening appears at the Fermi wave vector with a periodic modulation of the electronic density. From a structural point of view, the metal-insulator transition is associated with a Periodic Lattice Distortion (PLD) of the crystal lattice at the same Fermi wave vector 2k$_F$.

The CDW phase is highly sensitive to many external excitations such as temperature\cite{Moudden_prl90}, ultra-short laser pulses\cite{Singer2016,Jacques2016,Zong2019b,Yusupov2008,Gonzales2022,schm08,Kogar2020} or pressure\cite{Zocco_PRB2015}. CDW materials are also  strongly sensitive to chemical pressures\cite{Ru2008} or tensile stress \cite{straquadine_PRX2022,gallofrantz2023chargedensitywaves}.
Another property of incommensurate CDW is its ability to {\it slide}. Under an external applied field, CDW systems may display an excess of current. This charge transport  is highly specific because it is a pulsed and collective mode which occurs at the macroscopic scale and leads to electronic noise oscillations, commonly referred to as Narrow-Band Noise\cite{Monceau_PRB82,Fleming_prl1979}. 

The sliding phenomenon is well observed by transport measurements but it is also observed by diffraction since the transport of charges is associated to deformations of the PLD. Two types of deformations were observed by diffraction, along and tranverse to the applied field. A change of the $2k_F$ norm was observed at the threshold current close to electrical contacts \cite{PhysRevLett.70.845,PhysRevLett.80.5631}, while shear was observed through a change in the transverse width of the satellite peak\cite{bellec2019evidence,lebolloch2025}. All these measurements were performed in the NbSe$_3$ model system. In the TbTe$_3$ system studied here, the situation is more complex. TbTe$_3$ stabilizes also an incommensurate CDW which is able to slide\cite{Sinchenko2012} and the slding state is associated with a PLD deformation\cite{Lebolloch_prb2016}. However, deformations in TbTe$_3$, although very small in amplitude, include the three directions of the lattice and mainly the tranverse direction along b$^*$.

\begin{figure}[]
\centering
\includegraphics[width=\linewidth]{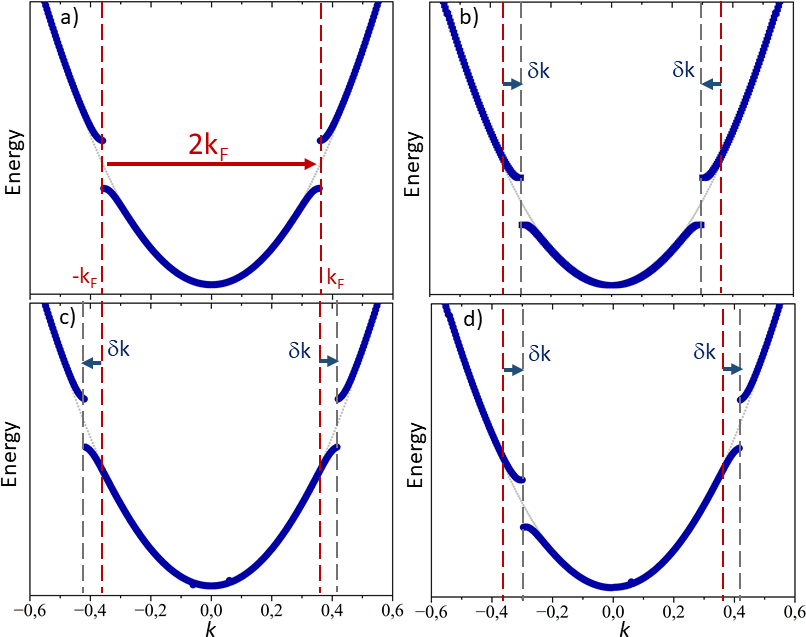}
  \caption{Scheme of displaced band structure of 1D CDW in the case of position and time-dependent phase. a) Gap opening at $\pm$ k$_F$ with constant phase with 2k$_F$=5/7. The gray curve correspond to the electronic dispersion of the metallic high temperature phase. b) Shift of the Fermi wave vector with $\delta k=\pm 1/2 \delta \phi(x)/\delta x$ in the case of contraction of the CDW wavelength  ($\delta \phi(x)/\delta x>0$) and c) in the case of dilatation of the CDW wavelength ($\delta \phi(x)/\delta x<0$).  d) Assymetric gap openning in the case of time-dependent phase with $\delta k=1/(2\hbar v_F)\delta \phi(t)/\delta t$.}
  \label{fig1}
\end{figure}

\begin{figure}[ht]
\centering 
  \includegraphics[width=\linewidth]{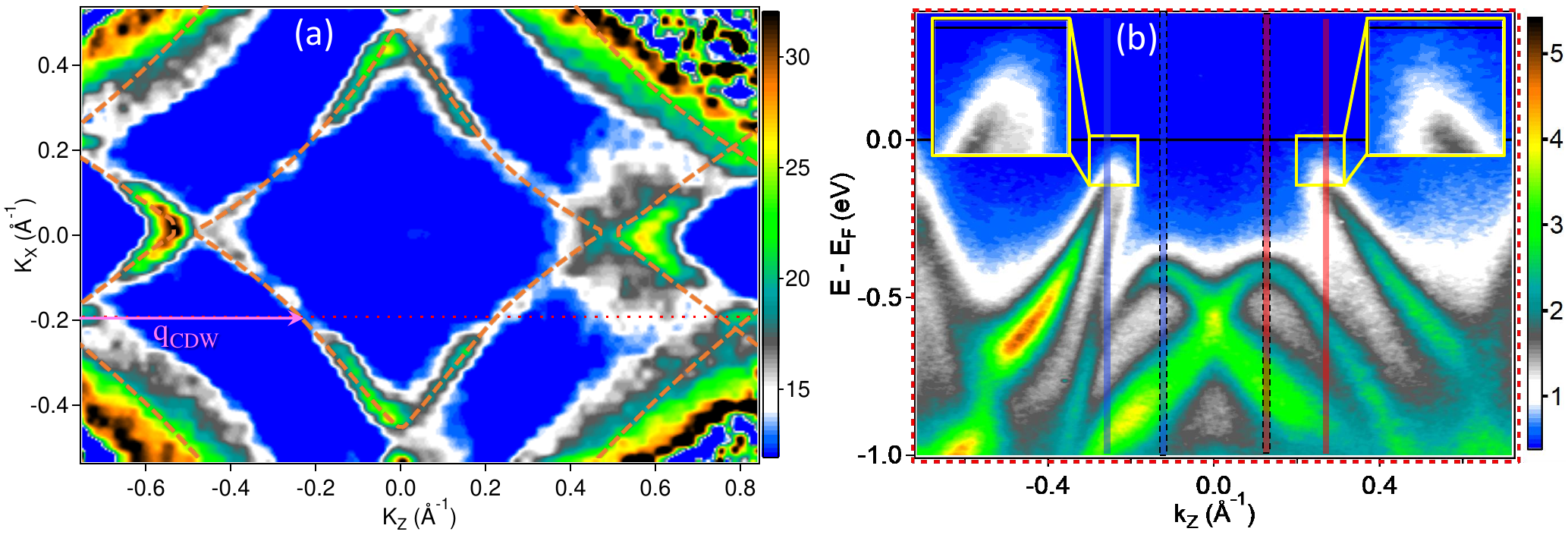}
  \caption{(a) Fermi surface of TbTe\textsubscript{3} at RT and without current. The dashed  lines are a guide to the eye of the Fermi surface sheets. The nesting vector $q_{CDW}$ is shown as well as red dashed line located at $k_x = -0.192$ {\AA$^{-1}$}  where the band dispersion has been measured. (b) Band dispersion along the red dotted line in with the CDW  gap. Zooms near E\textsubscript{F} are shown to better appreciate the gap. The CDW band gap shift versus applied current and a reference ungapped band have been analyzed in profiles with full and dashed lines respectively.}
  \label{fig2}
 \end{figure}

\begin{figure}[hb]
\centering
  \includegraphics[width=\linewidth]{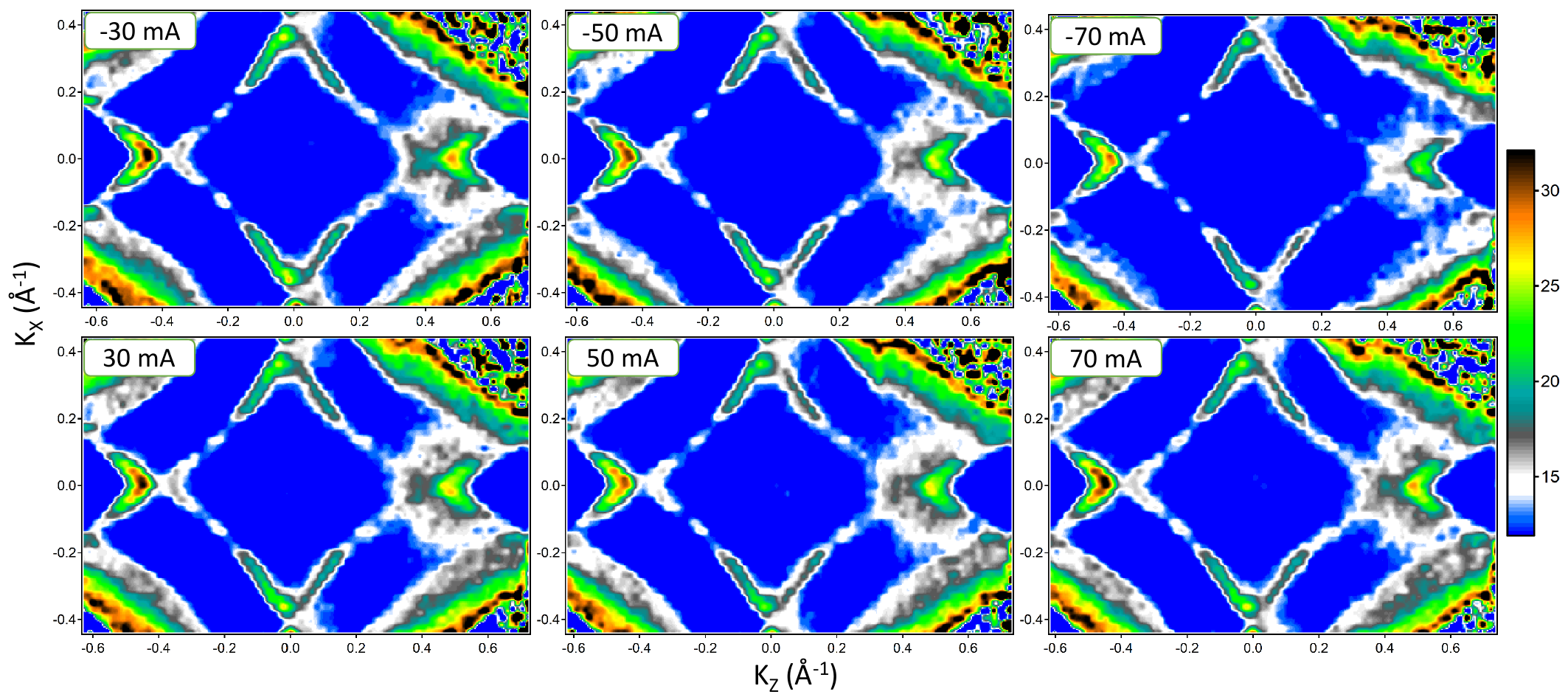}
  \caption{Fermi surfaces of TbTe\textsubscript{3} for different applied currents (integration window across the Fermi energy of 20 meV).}
  \label{fig5}
\end{figure}

 \begin{figure*}
\centering
\includegraphics[width=0.8\linewidth]{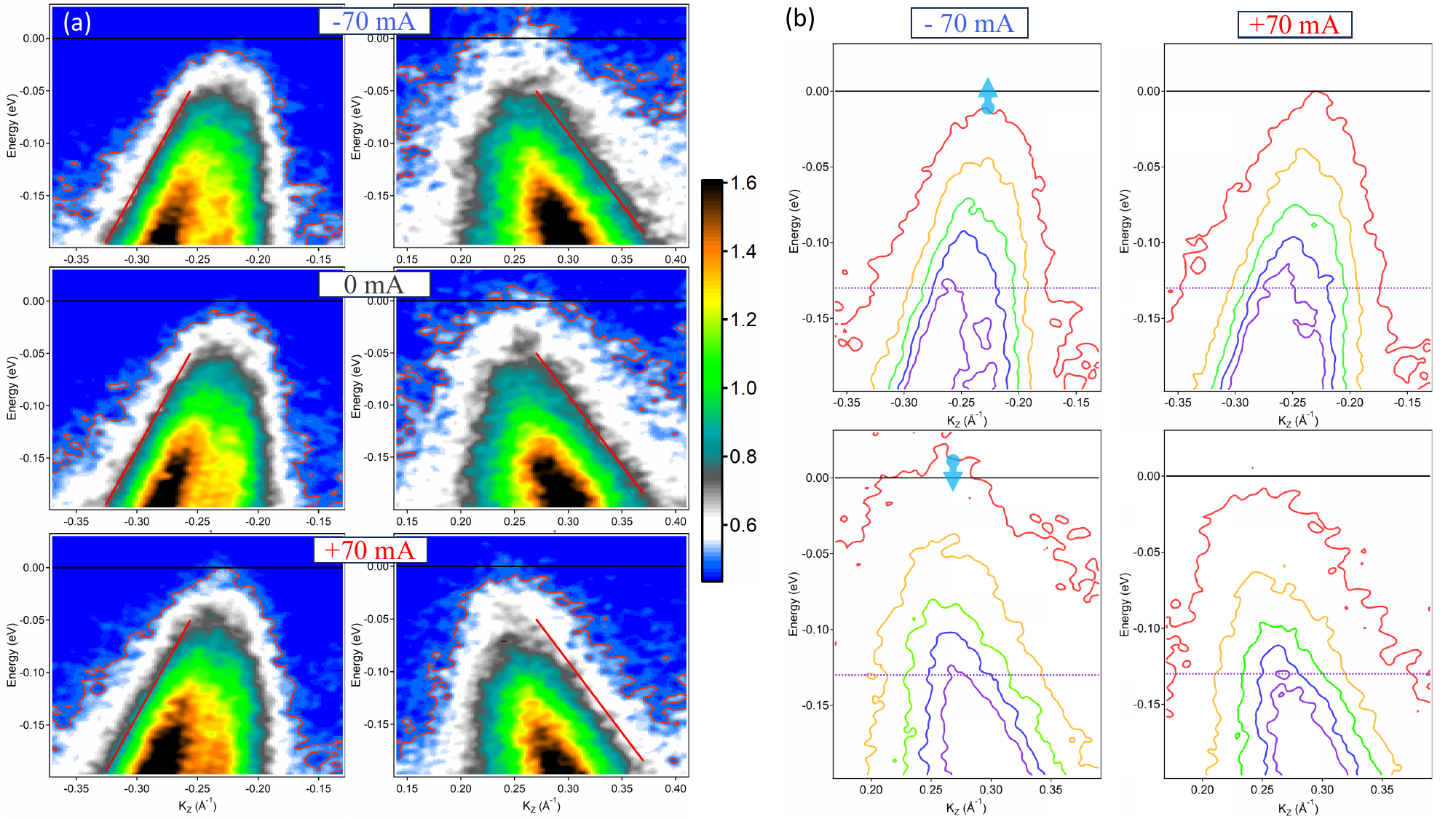}
\caption{(a) Detail at E\textsubscript{F} of the band dispersion at $k_x = -0.192$ {\AA$^{-1}$} for different applied currents ($-70$ mA (top row), $0$ mA (middle row), $+70$ mA(bottom row)). Since the intensity for positive and negative $k$ is different due to photoemission matrix elements, peak intensities have been normalized to the peak intensity at zero current. Red straight lines drawn at the band edges serve as guides to the eye, emphasising the horizontal (wavevector) shift of the bands. (b) Intensity contour plots for the left band (k$<$0, top row) and right band (k$>$0, bottom row). Positive current induces an upward shift (downward) on the left (right) band. Negative current induces an opposite behavior.}
\label{fig3}
\end{figure*}

These deformations and the excess of current are well described by a time and space dependent phase $\phi(\vec r,t)$ of the periodic modulation $\cos(2 k_F x+\phi(\vec r,t))$.
The additional current $j$ is related to temporal derivative $\frac{\delta\phi(t)}{\delta t}$, while CDW deformations are related to the spatial one, including the longitudinal ($\delta\phi/\delta x$) and the transverse ($\delta \phi/\delta y$) deformation. In both cases, PLD deformations and the excess current alter the band structure. However, the band deformation is profoundly different in the two cases: the former changes the k$_F$ value while preserving k-symmetry, whereas the latter breaks k-space symmetry (see Fig.1).
 In the case of a longitudinal deformation $\delta\phi/\delta x$, corresponding to an expansion or a contraction of the CDW period with a change in $2k_F$ and therefore to a change of the band filling (see Fig. 1b-c). In that case, the modification of the dispersion curve preserves the  symmetry in k-space since any spatial deformation do not favour one direction over another. On the other hand, the existence of an additional current $j$ across the crystal essentially breaks the band symmetry by favouring electrons along the direction of the applied field (see Fig. 1d).  The band structure is shifted of $\delta k$ due to the additional momentum $\delta p=\hbar \delta k=m^* v_s$, where $v_s$ is the sliding velocity,

$$v_s=\frac{1}{2k_F}\frac{\delta\phi}{\delta t}$$

Furthermore, the band structure is also shifted in energy. In the simple case of free like electron dispersion  curve ($\epsilon=\hbar^2k^2/2m^*$), the shift in energy is \cite{Allenderbardeen_1974}:  
$$\delta \epsilon=\hbar k v_s.$$
In fact, the shift in energy depends of the Fermi velocity $v_F$. The case where the curvature of the band clsoe to the gap openning is taken into account is treated in \cite{gruner1988}.

\begin{figure}[h]
\centering
\includegraphics[width=\linewidth]{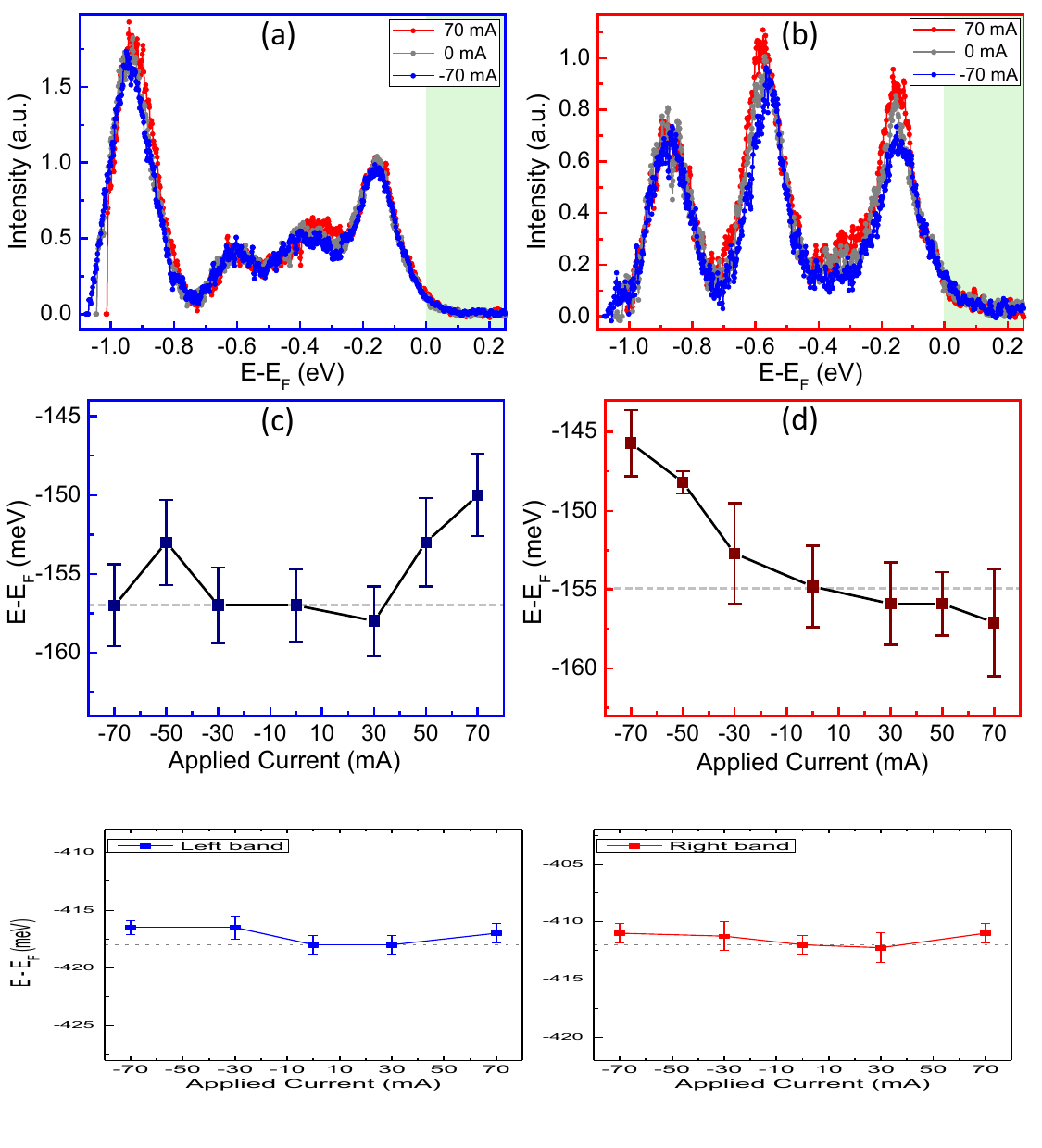}
\caption{Spectra evolution with applied current. (a, b) EDCs for 0 and $\pm 70$ mA applied currents. Spectra have been obtained from $k$ integration in the regions delimited by blue (left) and red (right) lines in Fig. \ref{fig2}(b). (c, d) Evolution as a function of the applied current of the closest peak to the Fermi level (around $\sim$-0.15 eV) for the left and right integrated line profiles, respectively. Error bars represent the standard deviation of the peak position. (e,f) Evolution as a function of the applied current of the semiconducting bands for the left and right integrated line dashed profiles in fig. 2. Error bars represent the standard deviation of the peak position.}
\label{fig4}
\end{figure}
In any case, if the signature of the siding state through the band structure should be present, it should be weak due to the orders of magnitude involved, making observation difficult by ARPES. However, in an attempt to observe it, the TbTe$_3$ system has been chosen because it displays a large gap ($\delta\approx$ 230 meV)\cite{Brouet2008} as well as sliding properties\cite{Sinchenko2012}. In this work, we will present experimental measurements compatible with the onset of sliding CDW.

TbTe$_3$ is a layered quasi-2D crystal stacked along the $b$-crystal axis which undergoes a CDW transition at 335 K with 2k$_F$=5/7 [$c^*$] along the $c$-axis.  The TbTe$_3$ sample was studied by ARPES at the ANTARES beamline of SOLEIL synchrotron at a photon energy of 30 eV, at room temperature and low pressure $\sim5 \times10^{-10}$ mbar. The sample was cleaved inside the ultra-high vacuum (UHV) chamber. The analyser slit was parallel to the $c$-axis, first determined from inhouse X-ray diffraction measurements. The beam size was 100$\mu m^2$ and located far from the electrodes to avoid any perturbation. The threshold current (I$_s$=30mA) has been checked before and after the experiment and the Fermi surface and the electronic band dispersions has been measured versus applied currents, below and above the threshold current. Fermi Surfaces (FS) are determined by integrating the photoemission signal in a 20 meV integration window across E$_F$. A Shirley background has been subtracted from the line profiles drawn on the Energy Distribution Curves (EDCs) for quantitative analysis. The energy reference was selected the Fermi level determined from the inflection point of the Fermi step.

The Fermi Surface is presented in Fig. \ref{fig2}a without current. The investigations to observe tiny band structure modifications have been studied at $k_x = -0.192$ {\AA$^{-1}$}  (red dotted line in Fig. \ref{fig2}a) located far enough from intense bands. The corresponding surface sheet  is shown without current in Fig. \ref{fig2}(b). A certain asymmetry of the bands is observed between k$>$0 and k$<$0 because of photoemission matrix elements and a slight sample misorientation during measurements. Insets in Fig. 2(b) are shown for a better visibility of the CDW-induced band gaps. 

 The Fermi surface at different applied currents below and  threshold current I$_s$, negative and positive, are shown in Fig.3. As expected, the overall FS exhibits only minor changes. 
Fig. 4 thus shows close-ups of the gapped regions for 0 and $\pm 70$ mA currents in order to appreciate if  there is any displacement of the band onset with respect to E\textsubscript{F}. The energy band offset shift can be appreciated at the  intensity contour plots in panel $b$ and the k-shift in panel $a$ by comparing to the red line highlighting the band edge at zero current. It can be appreciated that (i) positive currents promote an upward energy shift at the left band (k$<$0) and an opposite downward shift in the right band (k$>$0), (ii) For negative currents, the opposite behavior is observed and (iii) when the right band shifts to higher binding energies at positive current, it exhibits a concomitant $\Delta k <0$ shift. All these changes are small and close to our experimental resolution but seem to be compatible with expected results. 

In order to quantify the energy shift under applied current, Fig. \ref{fig4}(a-b) shows the EDCs indicated by the lines in Fig.  \ref{fig2}(b) at $k_z = \pm0.264$ {\AA$^{-1}$. The fitted peak energies as a function of applied current are shown in Figures \ref{fig4}(c) and \ref{fig4}(d). At $k<0$ (left side of the $\Gamma$ point), the two visible bands in the gap region make difficult a very precise determination. At k$>$0, the intensity of one of the two bands is significantly reduced due to the photoemission matrix elements. However, a small but systematic energy shift is observed, where the band shifts to higher binding energies for positive current  to lower binding energies for negative current. The estimated energy shift between the $+70$ mA and $-70$ mA is $\sim$12 meV. 

A global energy shift is induced by the applied electric field. For a comparison between currents, the EDCs have been rescaled from the inflection point of the Fermi step. For this, we have analyzed the energy shift of a reference band not affected by a CDW gap (Fig. 4 (e-f) corresponding to the dashed lines in fig. 2b). Fig. 4e-f shows the energy shift of the reference bands (k$<$0 and k$>$0) as a function of the applied current. The energy shift is $\sim 2$ meV and much smaller than the energy shift of the CDW gapped band.  This shows that the electric field is not affecting all the bands but preferable the bands involved in the CDW formation.


In conclusion, we have investigated the deformation of the band structure under electrical current predicted by the Fr\"ohlich theory for a Charge Density Wave (CDW) phase through an analysis of the Fermi surface and the electronic band dispersion of TbTe\textsubscript{3} measured by ARPES. 
A slight energy shifts of the CDW band gaps as a function of applied current were observed, which reverse when the current is reversed. This effect has to be confirmed but seems to be in agreement with expected behavior of band structure in the slding CDW regime.

\medskip
\textbf{Acknowledgements} \par 
The authors would like to thank Véronique Brouet and Gilles Abramovici for useful discussions.

\medskip

\bibliography{biblio_lcls_AT}

\end{document}